# Microfabricated high-finesse optical cavity with open access and small volume

M. Trupke, E. A. Hinds, S. Eriksson, and E. A. Curtis
*Centre for Cold Matter, Imperial College, Prince Consort Road, London, SW7 2BW, United Kingdom*

Z. Moktadir, E. Kukharenka, and M. Kraft
*School of Electronics and Computer Science, Southampton University, Southampton, SO17 1BJ, United Kingdom*

We present a novel microfabricated optical cavity, which combines a very small mode volume with high finesse. In contrast to other micro-resonators, such as microspheres, the structure we have built gives atoms and molecules direct access to the high-intensity part of the field mode, enabling them to interact strongly with photons in the cavity for the purposes of detection and quantum-coherent manipulation. Light couples directly in and out of the resonator through an optical fibre, avoiding the need for sensitive coupling optics. This renders the cavity particularly attractive as a component of a lab-on-a-chip, and as a node in a quantum network.

High-finesse optical cavities are central to many techniques and devices in atomic physics,[1] optoelectronics,[2] chemistry,[3] and biosensing.[4] As well as selecting spectral and spatial distributions of the classical electromagnetic field, optical cavities make it possible to harness quantum effects for applications in quantum information science.[1,5] For example, it is possible to produce single photons on demand using atoms[6,7] or ions[8] inside a cavity and to create entanglement between those that share a cavity photon.[9,10,11] Similar ideas are being pursued with quantum dots.[12,13] Microscopic cavities are of particular interest[14] because small volume gives the photon a large field and because they offer the possibility of integration with micro-opto-electro-mechanical systems[15] and atom chips.[16,17,18] Here we present a simple and innovative method for fabricating microscopic, broadly-tuneable, high-finesse cavities. These have the significant new feature that their structure is open, giving an atom, molecule or quantum dot direct access to an antinode of the cavity mode. This structure is therefore ideally suited for detecting small numbers of particles,[19] and miniaturizing quantum devices based on strong dipole-cavity coupling. We have made high-finesse, open optical cavities that operate in a length range of approximately 20 $\mu$m to 200 $\mu$m. Each cavity is formed by a concave micro-mirror and the plane tip of an optical fibre, both coated for reflection, as illustrated in Figure 1(a).

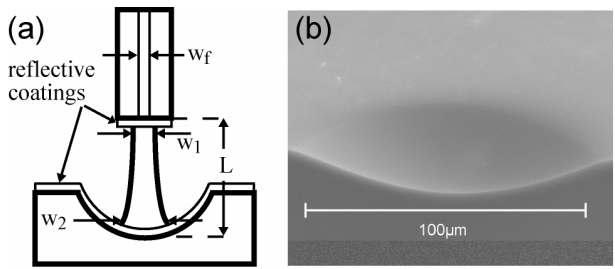

**Figure 1 (a)** Schematic diagram of the cavity. The light field drops to 1/e of its central value at radii $w_f$=2.65 $\mu$m in the fibre, $w_1=(\lambda/\pi)^{1/2}(LR-L^2)^{1/4}$ at the waist of the cavity mode and $w_2=(\lambda/\pi)(L/R)^{1/2}/w_1$ on the concave mirror. Here $L$ is the cavity length and $R$ is the radius of curvature of the lower mirror. **(b)** Scanning electron microscope image of an uncoated mirror template cleaved almost across the diameter.

Arrays of concave mirrors are fabricated in silicon by wet-etching isotropically through circular apertures in a lithographic mask using a mixture of HF and $HNO_3$ in acetic acid. The etch bath in which the wafer is immersed undergoes continuous agitation during the etching process, resulting in an approximately spherical surface profile, as shown in Figure 1(b). The etch rate and the final morphology of the silicon surface are highly dependent on the agitation and on the concentration of each component in the etchant.[20] Precise control over these factors gives us repeatable surface profiles in the silicon with 6 nm rms roughness. In our first experiment, gold is sputtered onto an array of mirror templates to form a layer 100 nm thick with a surface roughness of 10 nm. The plane mirror of the cavity is a dielectric multilayer, which we transfer from a custom-made substrate[21] to the cleaved tip of an optical fibre using index-matching epoxy glue.

We couple laser light of wavelength $\lambda = 780$ nm into the fibre and vary the separation $L$ between the two mirrors by moving the silicon wafer on a piezoelectric stack to view the cavity resonances as dips in the reflected intensity. The small length change of the cavity between adjacent radial modes is well approximated by[22] $(\lambda/2\pi)\arccos[\sqrt{1-L/R}]$. This allows us to measure the radius of curvature $R \cong 185$ $\mu$m of the gold mirror surface. A small splitting of the radial modes indicates a slight astigmatism of the mirror that amounts to a ~4 $\mu$m difference between the maximum and minimum values of $R$. After some careful alignment, we obtain a reflection spectrum that is dominated by the lowest radial cavity mode, normally labelled (0,0), with successive longitudinal modes separated by length changes of $\Delta L \cong \lambda/2$. Figure 2(a) shows an example of the reflected power spectrum measured over two such neighbouring modes. We obtain the finesse $F$ of the cavity from the ratio of $\Delta L$ to the full width at half maximum $\delta L$ of the resonances.

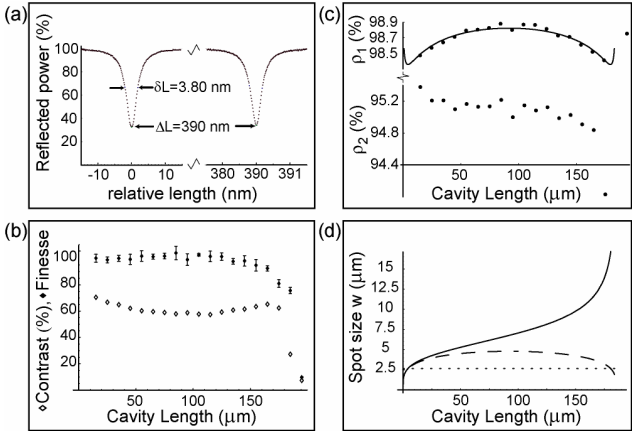
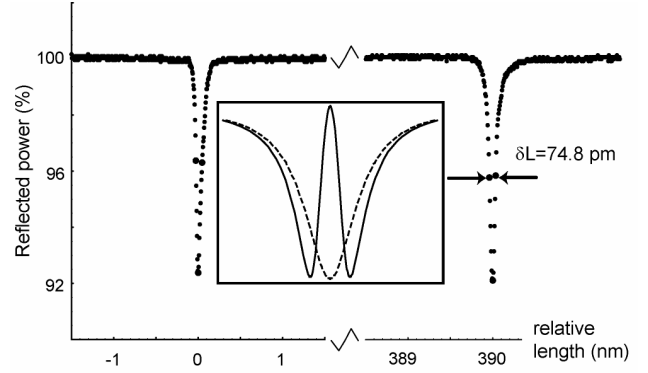

**Figure 2** Gold cavity data. **(a)** Successive longitudinal resonances in the (0,0) transverse mode of the micro-cavity. **(b)** Measurements versus cavity length of contrast and finesse. Error bars are the standard deviations of five independent measurements. **(c)** Effective reflectivities of the mirrors. Circles: measured values. Line: Theory for the plane mirror end. **(d)** Spot sizes on the curved mirror $w_2$ (solid line), plane mirror $w_1$ (dashed), and fibre $w_f$ (dotted).

This ratio is related to effective reflectivities $\rho_1$ and $\rho_2$ of the two mirrors, discussed below, by the formula[22]

$$F = \frac{\pi \sqrt[4]{\rho_1\rho_2}}{1-\sqrt{\rho_1\rho_2}} \approx \frac{\Delta L}{\delta L}. \quad (1)$$

From the same spectrum we also obtain the fringe contrast $C$, which we define as

$$C = 1 - \frac{I_{min}}{I_{max}} \approx 1 - \left(\frac{\sqrt{\rho_1}-\sqrt{\rho_2}}{1-\sqrt{\rho_1\rho_2}}\right)^2. \quad (2)$$

At the high reflectivities of our mirrors, both approximations are good enough to be considered equalities. The values of $F$ and $C$, measured for a variety of cavity lengths, are plotted in Figure 2(b) as filled and open diamonds respectively. The contrast exhibits a slight minimum near $L = R/2$, while the finesse is roughly constant at ~100, corresponding to a cavity $Q$-factor $2LF/\lambda$ of up to $4\times10^4$. Both drop dramatically near the limit of stability at $L = R$. Figure 2(c) shows the corresponding values of $\rho_1$ and $\rho_2$ derived from these measurements using equations (1) and (2) at each cavity length. We find that $\rho_1$ describing the plane mirror involves two factors: the intrinsic reflectivity $\rho$ of the dielectric stack, and the square of the overlap integral $\eta = 2w_f w_1/(w_f^2+w_1^2)$ that projects the field of the fibre mode onto the cavity mode. Thus we take $\rho_1 = 1-\eta^2(1-\rho)$, which varies with cavity length through the variation of $w_1$ given in the caption to Figure 1. The solid line in Figure 2(c) shows that this simple theory fits $\rho_1$ well[*] with $\rho = 98.4\%$, which is consistent with the manufacturers' specification of $98\% < \rho < 99\%$.

---

[*] A dip can be seen at the confocal length when $L = R/2$ because all the even radial modes are degenerate here. A simple extension of the theory that sums over these cavity modes is able to reproduce the dip.

**Figure 3** Cavity data with dielectric coatings. Main curve: successive longitudinal resonances in the (0,0) transverse mode at a cavity length of 25 $\mu$m. Inset: calculated splitting of resonance fringe (dashed) when one atom is inserted into the cavity (solid line).

The effective reflectivity $\rho_2$ of the gold mirror shows a slight decline with increasing cavity length, then plummets as the cavity approaches $L = R$. This shows that the curved mirror is the one responsible for the loss of finesse at large cavity lengths, presumably because of the large increase in spot size $w_2$, plotted in Figure 2(d).

Our primary interest is to use this cavity for detecting and coherently manipulating the state of single atoms trapped on an atom chip. We therefore consider the coupling between the cavity and a rubidium atom, for which the resonant dipole strength at 780 nm is characterised by the spontaneous decay rate $\Gamma = 4\times10^7$ s$^{-1}$. This is to be compared with the single photon Rabi frequency $g$, which describes the atom-cavity coupling strength at an antinode of the field, and with the damping rate $\kappa$ of the cavity field. These can be written as[23]

$$g = \sqrt{\frac{3\lambda^2}{\pi^2 w_1^2}\frac{c\Gamma}{L}} \quad ; \quad \kappa \approx \frac{\pi}{F}\frac{c}{2L}. \quad (3)$$

Because the cavity is short and the waist size of the cavity mode is only 4-5 $\mu$m, as shown in Figure 2(d), $g$ is much greater than $\Gamma$, for example $g = 9\times10^8$ s$^{-1}$ at a cavity length of 150 $\mu$m.

Now let us suppose that a single atom is introduced into the cavity and that the cavity is tuned to the atomic frequency. In that case, standard cavity QED theory[23] predicts that the amount of light absorbed at line centre will go from $I_{max}C$ to $I_{max}C/(2g^2/\kappa\Gamma+1)^2$. For the spectrum shown in Figure 2(a) this means that the fraction of light reflected at resonance will rise from 0.3 to 0.9, a dramatic increase that should allow the detection of a single atom with ease.[19]

Encouraged by this result, we built a higher-finesse microcavity by using coatings designed for 99.99% reflectivity on both mirrors. We measured with an atomic force microscope that the roughness $\sigma$ on the surface of the curved mirror was 2 nm, reducing the expected mirror reflectivity to approximately 99.9% through the scattering factor[24] $\exp[-(4\pi\sigma/\lambda)^2]$. When this cavity is 15 $\mu$m long, we obtain a finesse of 6000, which is consistent with these reflectivities. The finesse drops to 4000 at a cavity length of 105 $\mu$m, but even so, the cavity $Q$ is large at over $10^6$. A typical example of two successive resonances in the (0,0) transverse mode is shown in Figure 3, where the cavity length is 25 $\mu$m and the finesse is 5200. The corresponding cavity QED parameters approach the strong coupling regime, with $g = 2.3 \times 10^9$ s$^{-1}$ and $\kappa = 3.6 \times 10^9$ s$^{-1}$. Here $g^2/\kappa\Gamma = 39$. This atom-cavity coupling is large enough for a single atom to produce the substantial splitting of the resonance shown inset in Figure 3. This brings us into a new regime for open micro-resonators where schemes for the deterministic creation of single photons become feasible.[25] With suitably different reflectivities at the two mirrors we can ensure that the photons created in this cavity are coupled with high probability into the fibre. This forms an ideal basis for transporting quantum information throughout a network. In this regime, the cavity is also suitable for cooling[26,27] or entangling[9,10,11] atoms or molecules that are guided into it. With some smoothing of the substrate surface it will be possible to make $g$ significantly larger than $\kappa$. This could be achieved, for example, by plasma etching,[28] or by coating the silicon with a borosilicate layer which can be smoothed by thermal reflow.[29]

The micro-mirrors used in these resonators are immediately applicable to a wide variety of devices as their manufacture only involves standard silicon etching and optical coating techniques. The repeatability of the etching and coating means that patterns of many cavity mirrors can be produced rapidly and consistently. They can easily be combined with other silicon-based devices such as those in microfluidics, which are just starting to employ optical tools,[30] and in other lab-on-a-chip applications. In the specific context of atom chips, these cavities can readily be incorporated into the magnetic traps and guides available on current devices,[16,17,18] which are precise enough to allow the transport of atoms in and out of the cavity at will. The performance already achieved with these cavities, and the possibility of higher finesse still, show that they are ideal candidates for integrated cavity QED and quantum information processing applications, potentially in large arrays. Dielectric cavities, such as microtoroids and microspheres,[14] can have higher $Q$-factors than we have demonstrated. However, they are not as straightforward to use because both the atom and the external light field must couple to the cavity mode through its evanescent tail outside the dielectric. By contrast this structure offers direct access to the interior of the cavity and also couples the light in and out in a simple way through an integrated fibre. These two important differences provide an elegant solution to the problem of achieving atom-cavity coupling on a chip.

**Acknowledgements:** We acknowledge expert advice from Bruce Klappauf and Steve Helsby, and the expert technical assistance of Jon Dyne and Bandu Ratnasekara. This collaboration is supported by the UK Basic Technology programme. Additional support is provided by the Quantum Information IRC of the UK Research Councils, the Royal Society, and the European Commission's FASTNET and Atom Chips training network.